\documentclass{cernepprep} 
\usepackage{cernunits,cernchemsym,heppennames2}
\usepackage{cite,fancyvrb}

\usepackage{graphicx}
\usepackage{dcolumn}
\usepackage{bm}
\usepackage[utf8]{inputenc}
\usepackage[T1]{fontenc}
\usepackage{mathptmx}
\usepackage{etoolbox}
\usepackage{amsmath}
\usepackage{textgreek}

\begin{document}

\begin{titlepage}
\EPnumber{2023-102}
\EPdate{\today}
\DEFCOL{CDS-Library}

\title{SDR, EVC, and SDREVC: Limitations and Extensions}

\begin{Authlist}
 
E.~D.~Hunter$^{a}$\thanks{Electronic mail: eric.david.hunter@cern.ch},
C.~Amsler$^{a}$, 
H.~Breuker$^{b}$,  
M.~Bumbar$^{a}$, 
S. Chesnevskaya$^{a}$, 
G. Costantini$^{c,d}$,
R. Ferragut$^{e,f}$, 
M. Giammarchi$^{f}$, 
A. Gligorova$^{a}$,  
G. Gosta$^{c,d}$, 
H.~Higaki$^{g}$, 
C.~Killian$^{a}$, 
V.~Kraxberger$^{a,h}$, 
N.~Kuroda$^{i}$, 
A.~Lanz$^{a,h}$,
M.~Leali$^{c,d}$,
G.~Maero$^{f,j}$, 
C.~Mal\-bru\-not$^{k}$\thanks{present address: TRIUMF, Vancouver, Canada}, 
V.~Mascagna$^{c,d}$, 
Y.~Matsuda$^{i}$,  
V.~M{\"a}ckel$^{a}$\thanks{present address: INFICON GmbH, K\"oln},
S.~Migliorati$^{c,d}$, 
D.~J.~Murtagh$^{a}$, 
A.~Nanda$^{a,h}$,  
L.~Nowak$^{a,h,k}$, 
F.~Parnefjord~Gustafsson$^{a}$, 
S.~Rhein\-frank$^{a}$, 
M.~Romé$^{f,j}$, 
M.~C.~Simon$^{a}$, 
M.~Tajima$^{l}$,  
V. Toso$^{e,f}$, 
S.~Ulmer$^{b}$,  
L.~Venturelli$^{c,d}$,
A.~Weiser$^{a,h}$, 
E.~Widmann$^{a}$, 
Y.~Yamazaki$^{b}$,  
J.~Zmeskal$^{a}$\\ 

\end{Authlist}

\Collaboration{(The ASACUSA-Cusp Collaboration)}

$^{a}$Stefan Meyer Institute for Subatomic Physics, Austrian Academy of Sciences, Vienna, Austria, 
$^{b}$Ulmer Fundamental Symmetries Laboratory, RIKEN, Saitama, Japan,
$^{c}$Dipartimento di Ingegneria dell'In\-formazione, Universit\`a degli Studi di Brescia, Brescia, Italy
$^{d}$INFN sez. Pavia, Pavia, Italy,
$^{e}$L-NESS and Department of Physics, Politecnico di Milano, Como, Italy,
$^{f}$INFN sez. Milano, Milan, Italy,
$^{g}$Graduate School of Advanced Science and Engineering, Hiroshima University, Hiroshima, Japan,
$^{h}$University of Vienna, Vienna Doctoral School in Physics, Vienna, Austria,
$^{i}$Institute of Physics, Graduate School of Arts and Sciences, University of Tokyo, Tokyo, Japan,
$^{j}$Dipartimento di Fisica, Universit\`a degli Studi di Milano, Milan, Italy
$^{k}$Experimental Physics Department, CERN, Geneva, Switzerland,
$^{l}$RIKEN Nishina Center for Accelerator-Based Science, Saitama, Japan,

\begin{abstract}
Methods for reducing the radius, temperature, and space charge of nonneutral plasma are usually reported for conditions which approximate an ideal Penning Malmberg trap. Here we show that (1) similar methods are still effective under surprisingly adverse circumstances: we perform SDR and SDREVC in a strong magnetic mirror field using only 3 out of 4 rotating wall petals. In addition, we demonstrate (2) an alternative to SDREVC, using e-kick instead of EVC and (3) an upper limit for how much plasma can be cooled to $T<20\,\mathrm{K}$ using EVC. This limit depends on the space charge, not on the number of particles or the plasma density.
\end{abstract}

\end{titlepage}

\section{Introduction}
\label{sec:intro}
Fine control over plasma parameters is beneficial for experiments that use destructive diagnostics\cite{hurs:16, ahma:17, eve:18}, have long cycle times\cite{sing:21, amsl:21, blum:22}, or struggle for limited resources like antimatter\cite{hori:13}, rare isotopes\cite{auma:22}, or highly charged ions\cite{klug:08}. For many of these experiments, the density and temperature of the plasma are key parameters which must be both reproducible, for the reasons just stated, and optimal. In practice, the maximum plasma density and minimum plasma temperature are far from the theoretical optimum, and it remains an open question whether the observed bounds on density and temperature are fundamental or merely a sign that better methods are required. 

The density of an electron or positron plasma can be continuously tuned using the strong drive regime (SDR) rotating wall technique of Danielson and collaborators\cite{dani:07}. The plasma temperature may be reduced via evaporative cooling (EVC)\cite{andr:10}. When the plasma is cold enough that its temperature $T$ is less than its space charge $\phi_0$ ($k_B T \ll e\phi_0$), then EVC also reliably defines the latter. Combining these two tools yields a new one, SDREVC, which allows the plasma density, length, space charge, and temperature to be set simultaneously\cite{ahma:18}.

When it works, SDREVC takes as input ``any'' plasma (of sufficiently many particles) and produces a user-defined, invariable (to ${\sim} 1\%$) final state\textemdash greatly facilitating optimization studies and systematic searches generally. But when \emph{does} it work? A priori, the experimenter cannot give an exhaustive list of the conditions necessary for these techniques to work as advertised. The main ingredient (SDR) is only heuristically described by theory. One must be content to find, empirically, some range of parameters which give good results for a particular trap.

It is useful to know what limits exist and which assumptions are actually relevant to the process. We explore these questions in experiments on electron plasma held in ASACUSA's Cusp Trap\cite{kuro:17}. Section~\ref{sec:app} reviews aspects of the control and diagnostic systems relevant for this work. Each technique is addressed in its own section: SDR in Section~\ref{sec:SDR}, SDREVC in Section~\ref{sec:SDREVC}, and EVC in Section~\ref{sec:EVC}. Section~\ref{sec:conclu} contains a short discussion of the results, summarizing what can be used for testing a potential theory, and indicating where practical extensions may be possible.

\section{Apparatus}
\label{sec:app}

We use a Penning Malmberg trap\cite{malm:80} with inner diameter $34\,\mathrm{mm}$ and magnetic field $B\approx 2\,\mathrm{T}$. The entrance and exit to the trap are screened by $80\%$ transparent copper meshes so that microwave radiation from the plasma cannot escape the $6\,\mathrm{K}$ cryogenic ultra high vacuum region\cite{amsl:22}. A typical plasma contains $N\sim 10^7$ electrons, with length $L_p\approx 10\,\mathrm{cm}$, radius $r_p\approx 1\,\mathrm{mm}$, and density $n\sim 10^8\,\mathrm{cm}^{-3}$. The plasma cools via cyclotron radiation to a steady state temperature $T\approx 30\,\mathrm{K}$ for $N\lesssim 4 \times 10^7$.

\subsection{Control}

\begin{figure}
    \centering
    \includegraphics[width=0.9\linewidth]{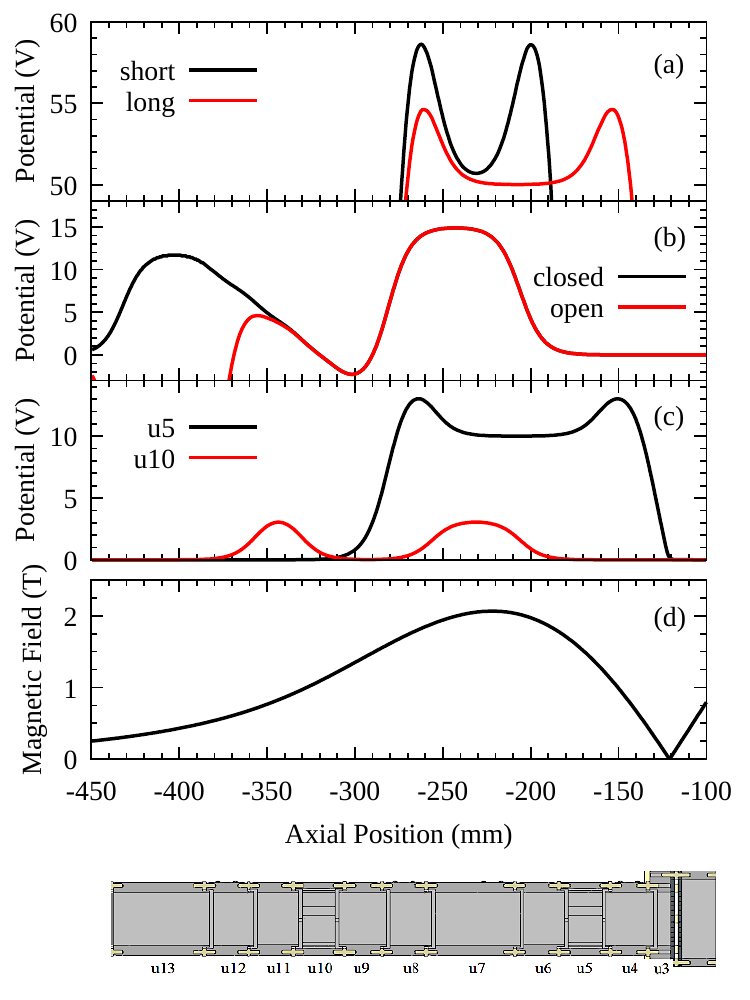}
    \caption{On-axis electrostatic potential for (a) EVC, (b) pulsed SDREVC, and (c) standard SDREVC. In this figure the voltages are multiplied by (-1) for clarity; they must be inverted for confining negatively charged particles. Magnetic field strength (d) is shown above a scaled drawing of the electrode stack.}
    \label{fig:pots}
\end{figure}

Two pairs of superconducting anti-Helmholtz coils produce the axial magnetic field shown in Fig.~\ref{fig:pots}(d). The strong gradients were designed to focus low field seeking antiatoms traveling to the right toward the spectroscopy beamline\cite{naga:14,naga:15}. Although the coils produce a strong field in three regions, it is difficult to move a plasma into the middle or right region because of the magnetic field nulls (cusp points) at $z=\pm 120\,\mathrm{mm}$. Plasma is loaded exclusively on the left side, either from an electron source (Nishinbo NJK1120A) or from other traps. 

One electrode, u13, must be pulsed for catching bunches of protons, antiprotons, and positrons from those traps (electrode locations are given at the bottom of Fig.~\ref{fig:pots}). Two electrodes, u5 and u10, are driven with radiofrequency for SDR and plasma or ion heating. The other electrodes (u12, u11, u9, u8, u7, u6, u4, u3) have cryogenic 2-pole RC filters mounted as close to the electrode as possible (${\sim}1\,\mathrm{cm}$ away). These filters attenuate noise with frequency $f>10\,\mathrm{kHz}$, which would heat the plasma. The pulsed electrode u13 has a similar filter, but the filter is bypassed by a pair of BAT54 Schottky diodes. 

The slower, nearly DC bias for the electrodes is supplied by low speed x15 high voltage amplifiers designed by J.~Fajans. These amplifiers exhibit superior stability (about $1\,\mathrm{mV}$ per year) and noise performance ($0.1\,\mathrm{mV_{rms}}$). The price for this performance is a rise time of almost $5\,\mathrm{ms}$. This limits the range of plasma dynamics which can be studied in our experiment; for instance, after EVC we must spend $10\,\mathrm{ms}$ changing the well shape before we can properly diagnose the plasma temperature. 

The rotating wall (RW) electrodes u5 and u10 have 4 azimuthal sectors each. Each sector is driven with a sine wave, phase shifted as $\{0^\mathrm{o},90^\mathrm{o},180^\mathrm{o},270^\mathrm{o}\}$. These signals are produced by the $\{0^\mathrm{o},90^\mathrm{o}\}$ outputs of a 2-channel arbitrary waveform generator (BK Precision BK4054B) followed by $180^\mathrm{o}$ phase splitters (Mini Circuits ZSCJ-2-2+). The signals then enter a filterbox mounted to the vacuum feedthru. The filterbox contains a low pass filter for DC biasing of the electrode ($RC=0.2\,\mathrm{ms}$) and high pass filters for the RW signals ($RC= 0.01\,\mathrm{ms}$). One sector of u10 is grounded inside the vacuum chamber (unintentionally); the resistance to ground is less than $4\,$\textOmega, and does not change when the experiment is cooled down.
The filter for u10 was modified such that a DC bias could still be applied to two opposing sectors, while the others are $50\,$\textOmega\ terminated. The AC inputs are essentially the same as for u5's filter. The impedance between RW sectors is approximately $50\,$\textOmega\ at frequencies above $f\sim 10\,\mathrm{kHz}$.

An HV pulser drives electrode u13 via a similar filter circuit with $RC=0.2\,\mathrm{ms}$. The pulser is similar to the one described in Ref.~\cite{chan:04}, with the addition of a transformer and relay to produce pulses of either polarity. 

\subsection{Diagnostics}

The temperature $T$, radius $r_p$, and number of electrons $N$ are determined by dumping the plasma out of the trap against a microchannel plate-phosphor screen detector (MCP)\cite{peur:93}. Since this operation destroys the plasma, only one of its properties can be accurately measured. Cycle-to-cycle reproducibility is therefore a basic assumption in this work. Using SDREVC we achieve $dN/N<1\%$ for any desired initial state. 

To get $T$, the plasma is released slowly (${\sim}1\,\mathrm{ms}$). The flux of escaping particles, measured by a SiPM, is combined with the time-dependent confinement potential to reconstruct the tail of a Maxwellian\cite{eggl:92}. On a graph of $\mathsf{log}\mathrm{[SiPM\ signal]}$ vs. $\mathrm{-[confinement]}$, the slope of the line that fits the rising edge is $e/k_BT$.

To get $r_p$ or $N$, the plasma is released quickly (${\sim}1\,\mathrm{\mu s}$) using the HV pulser described above. The radial profile is captured by a CMOS camera (Thorlabs CS165MU1) and fit to a modified Gaussian\cite{evan:16}. For $N$, the MCP is not biased; the front of the MCP is connected to an integrator which produces a voltage $V=GNe/C$, where $G=168$ is the amplifier gain and $C=1.09\,\mathrm{nF}$ is the combined capacitance of amplifier, cabling, and parasitics.

The plasma length $L_p$ and density $n$ can be derived from $r_p$, $N$, and the confining potentials. There exist several numerical plasma solvers for precise estimation of these quantities, but most of them assume a uniform magnetic field; see for example Ref.~\cite{pein:05}. Instead of developing a new code, here we simply compute $n\approx N/\pi r_p^2 L_p$, where $n$ and $r_p$, which actually vary with $B$, are understood to be an average over the length. $L_p$ is estimated as follows. We slowly release the plasma to measure its space charge $\phi_0$\textemdash this is the on-axis potential difference between well bottom and barrier when the first particles begin to escape. The plasma length at that instant is approximately the distance between the turning points of barely-confined particles. To get from this to the plasma length during RW compression, we note that the product of the plasma length and the space charge is roughly constant, i.e. $\phi_0 \propto N/L_p$ for a long plasma. The length in the RW well is thus chosen from a table of $L_p$ vs. $\phi_0 \times L_p$ for that well. 

\section{SDR}
\label{sec:SDR}

The rotating wall (RW) is used to change the radius, and hence the density, of the plasma\cite{ande:98}. An electric dipole field with an amplitude of order $1\,\mathrm{V/cm}$ revolves around the plasma in the plane perpendicular to the magnetic field. The field presumably perturbs the radial edge of the plasma; it may also cause the center of mass to oscillate (there is some speculation that the RW can recenter a plasma which began on a small diocotron orbit). In the strong drive regime (SDR) the plasma density $n$ evolves until the natural rotation rate of the plasma matches the RW frequency $f$\cite{dani:07}. 

\begin{figure}
    \centering
    \begin{minipage}{0.49\textwidth}
        \centering
        \vspace{0pt}
        \includegraphics[width=\linewidth]{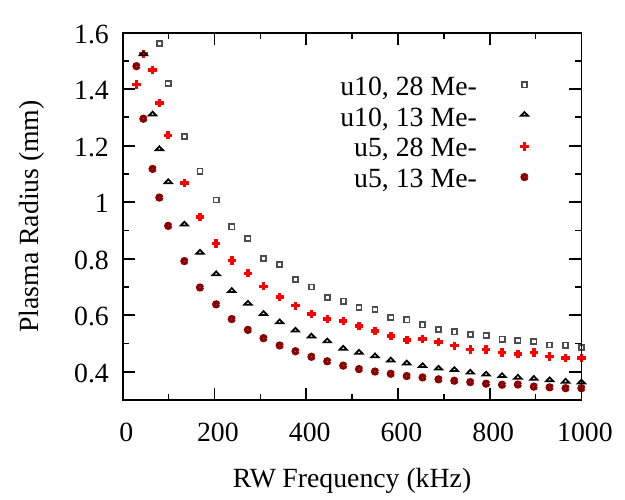} 
    \end{minipage}
    \hfill
    \begin{minipage}{0.49\textwidth}
        \centering
        \vspace{0pt}
        \includegraphics[width=\linewidth]{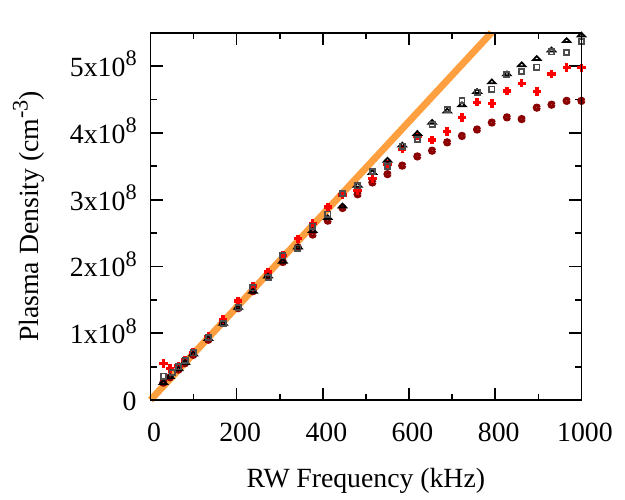} 
    \end{minipage}
    \caption{Control of the plasma density using SDR in u5 and u10, for $N=13\times 10^6$ or $N=28\times 10^6$. Left: plasma radius, measured as a function of applied RW frequency. Right: plasma density, evaluated as $n=N/(\pi r_p^2 L_p)$.}
    \label{fig:sdr}
\end{figure}

In this regime, a graph of plasma density vs. applied frequency should be a line starting from the origin with slope $n/f=4\pi\epsilon_0 B/e$\cite{ahma:18}. Practically, we search for some combination of on-axis potential, RW amplitude, and RW frequencies such that the linear relation is observed for the widest range of frequencies possible. Figure~\ref{fig:sdr} shows examples from our experiment. The data is obtained by first preparing $13\times 10^6$ or $28\times 10^6$ electrons via SDREVC at $200\,\mathrm{kHz}$, then moving to an SDR well (``u5'' and ``u10'' refer to wells similar to those shown in Fig.~\ref{fig:pots}(c)) and compressing for $10\,\mathrm{s}$ at a different frequency $f$ and higher RW amplitude (the RW waveforms are about $4\,\mathrm{V_{pp}}$ at the vacuum feedthrough). 

Compression in u5 is linear up to about $400\,\mathrm{kHz}$; for u10 the range is slightly higher. In general, the useful range of SDR is less than $1\,\mathrm{MHz}$ for our experiment. A similar limit was found in Ref.~\cite{ahma:18}. Danielson et al. suggested that SDR fails when the rotation rate is too close to one of the plasma's Trivelpiece-Gould modes, and that a stronger drive or a higher frequency can recover SDR in this case. We find, on the contrary, that the steady-state plasma temperature (while RW is still applied) increases monotonically with higher RW frequency, from $200\,\mathrm{K}$ at $500\,\mathrm{kHz}$ up to several $\mathrm{eV}$ at $3\,\mathrm{MHz}$. (Higher RW amplitude cannot be tested, but we do find that the amplitude can be decreased by a factor of 4 in u10 without changing the resulting radii.)

The density at a given frequency and magnetic field is the same for both values of $N$. For higher $N$, this means that the plasma radius (and length) is greater. For a longer plasma (u5 vs. u10), this means a smaller radius. These observations confirm our expectations for SDR compression. We conclude that the factor of two mirror field ratio does not compromise SDR in u5 or u10, nor does the fact that one sector of u10 is grounded. No particles are lost, no halo is produced, no diocotron is induced, and the steady-state temperature is low ($100\text{--}200\,\mathrm{K}$) in both wells over the linear (SDR) range of RW frequencies. 

We note, however, that our results do not reproduce the $B$ dependence expected from the relation $n/f=4\pi\epsilon_0 B/e$. First, the line in Fig.~\ref{fig:sdr} has a slope corresponding to $B=1\,\mathrm{T}$, which is a field value only reached at one extremity of the plasma; everywhere else the field is higher. Second, a given frequency should presumably give a denser plasma in u5 than in u10, because the average magnetic field in u10 is lower: ${<}B{>}=1.35\,\mathrm{T}$ in u10 and $1.87\,\mathrm{T}$ in u5. Instead, we find that the graph of $n\ \mathrm{vs.}\ f$ is nearly the same for both wells. The analysis may be compromised by other effects arising from the strong mirror field in the Cusp trap. A plasma spanning a large range of $B$ should still reach a rigid-rotor equilibrium defined by a single rotation rate, but it is by no means cylindrical\cite{faja:03}. A simple axial average ${<}B{>}$ does not reflect the true distribution of particles, partly because of magnetic mirroring. Moreover, the assumption that particles follow the field lines does not hold for nonneutral plasma equilibrium in a magnetic mirror. This latter effect would tend to make the u5 plasma expand more than the u10 plasma during transfer to the pulsed dump well used for imaging, albeit only by a factor of $5\%$ if we use the first approximation given in Ref.~\cite{faja:03}. 

\section{SDREVC}
\label{sec:SDREVC}

Starting with sufficiently many particles in any initial state, we use SDR to control the density and EVC to control the space charge\cite{ahma:18}. Since the plasma density and space charge completely define the cold plasma equilibrium for a given well shape, this operation also fixes the plasma radius and number of electrons. In this section, we will first show that SDREVC, like SDR, works as well in u5 as in u10, with its strong gradient and one grounded RW sector. We then demonstrate a different, pulsed protocol, where the space charge is reduced without EVC, resulting in similar or even improved reproducibility compared to standard SDREVC.

For this section we define ``any initial state'' to be any one of five starting points prepared using SDREVC in u5. That is, we begin with SDREVC with the parameters given in one of the rows of Table~\ref{tab:initial}. Then we change wells, frequency, and protocol for the final SDREVC to a fixed value. 

\begin{table}
  \begin{center}
  \begin{tabular}{|c|c|c|c|}
        \hline
      RW Freq. (kHz)  &  $\phi_0\,\mathrm{(V)}$  & No. e- ($10^6$) & Radius (mm)   \\
        \hline
       150  &   6.0 &   48.6  &   1.22\\
       200  &   5.5 &   41.1  &   0.99\\
       250  &   5.0 &   34.8  &   0.82\\
       300  &   4.5 &   29.0  &   0.70\\
       350  &   4.0 &   24.4  &   0.60\\
        \hline
  \end{tabular}
  \caption{Set of initial plasma preparations for testing different SDREVC routines.}
  \label{tab:initial}
  \end{center}
\end{table}

\subsection{Standard Protocol}

We implement SDREVC by starting in an SDR well and slowly reducing the confinement. In order to maintain a constant amount of overlap with the RW electrode as the space charge decreases, we reduce both sides of the confining potential simultaneously. The starting point of this operation is shown in Fig.~\ref{fig:pots}(c) for wells in u5 and u10. The excess charge is slowly released (EVC for $18\,\mathrm{s}$) while the density is locked by SDR at $200\,\mathrm{kHz}$. 

\begin{figure}
    \centering
    \begin{minipage}{0.49\textwidth}
        \centering
        \vspace{0pt}
        \includegraphics[width=\linewidth]{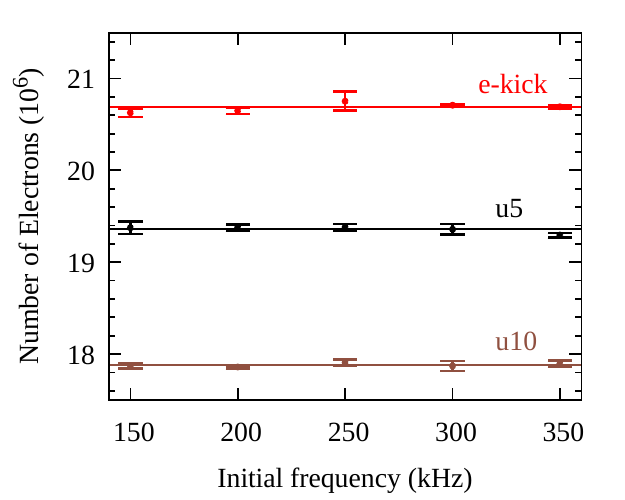} 
    \end{minipage}
    \hfill
    \begin{minipage}{0.49\textwidth}
        \centering
        \vspace{0pt}
        \includegraphics[width=\linewidth]{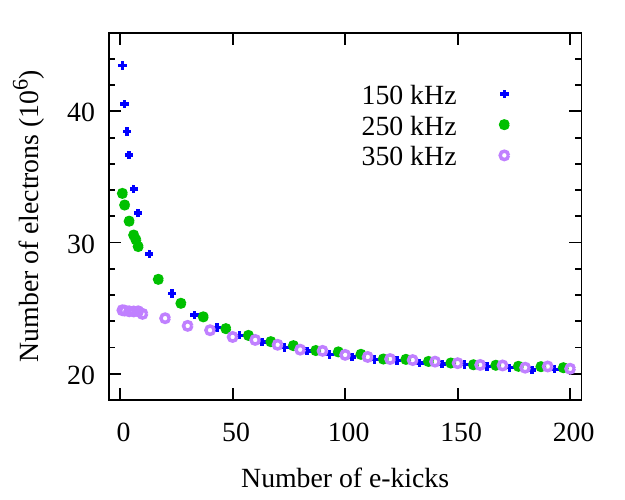} 
    \end{minipage}
    \caption{Control of the number of electrons using SDREVC and a variant with EVC replaced by e-kick. Left panel: Reproducibility of final number of electrons after SDREVC in u5 or u10, or the e-kick variant, for different initial plasma preparations (see Table~\ref{tab:initial}); data is reported as the mean of four measurements, with error bars showing $\pm 1$ standard deviation. Right panel: Convergence of the e-kick routine for three of these preparations; data points represent single measurements.}
    \label{fig:sdrevc}
\end{figure}

The final number of electrons is plotted in Fig.~\ref{fig:sdrevc} (left panel). Slightly different endpoints ($N=17.9$ and $19.4\times 10^6$) were chosen for the different SDREVC protocols so that the data would not overlap. The deviation from the endpoint value, as well as the standard deviation, is $1\%$ or less for any initial condition ($1\%$ is one minor tick in the figure). 

\subsection{Pulsed Protocol}

Instead of slowly reducing the confinement, we may use the pulsed electrode u13 to rapidly open and close the well such that only a fraction of the particles have time to escape. Such pulsing is sometimes referred to as an e-kick because electrons are thereby ``kicked out of'' a mixed electron-antiproton plasma without losing the antiprotons. We use a $65\,\mathrm{V}$, $50\,\mathrm{ns}$ HV pulse to briefly switch from the ``closed'' to ``open'' configurations shown in Fig.~\ref{fig:pots}(b). This is done once every $100\,\mathrm{ms}$, which means the plasma is in the ``closed'' well for $99.9999\%$ of the operation. 


The routines used to obtain the data in the left panel of Fig.~\ref{fig:sdrevc} employ 200 e-kicks, which amounts to $20\,\mathrm{s}$ of RW compression and pulsing. The right panel shows the evolution of $N$ as the number of e-kicks is varied. The convergence of different initial conditions to a common (exponential) curve is faster than the exponential convergence of the entire set to a final, minimum value of $N$.

The plasma temperature is $170\,\mathrm{K}$ at the end of standard SDREVC in u5, $130\,\mathrm{K}$ for the same in u10, and $400\,\mathrm{K}$ at the end of the pulsed protocol. Not surprisingly, the e-kicked plasma is hotter. The ramp-shaped well shown in Fig.~\ref{fig:pots}(b) gives better results with fewer e-kicks than a flat well. We assume this is either because the average magnetic field is higher (more cooling) than for a flat well spanning the same range in the axial direction, or because the ramp pushes the bulk of the plasma away from the HV pulses applied to u13 (less heating). We note that such a well does not allow for SDR compression; the graph of $n\ \mathrm{vs.}\ f$ is linear up to about $400\,\mathrm{kHz}$, but the intercept is slightly above the origin. The high level of reproducibility obtained with the pulsed protocol is remarkable, given these apparent disadvantages.

\section{EVC}
\label{sec:EVC}

Once a plasma has reached a low steady-state temperature, EVC is often used to reduce the temperature further. However, forced evaporative cooling does not always produce a lower temperature. In this section, we present data which suggests a link between the plasma space charge and the final temperature achievable using EVC.

\begin{figure}
    \centering
    \begin{minipage}{0.49\textwidth}
        \centering
        \vspace{0pt}
        \includegraphics[width=\linewidth]{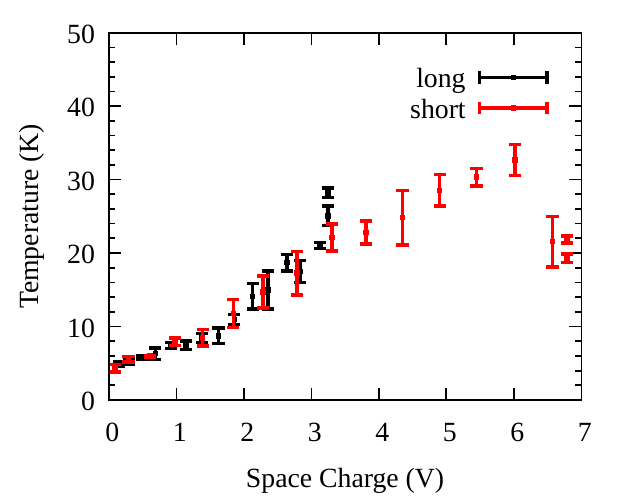} 
    \end{minipage}
    \hfill
    \begin{minipage}{0.49\textwidth}
        \centering
        \vspace{0pt}
        \includegraphics[width=\linewidth]{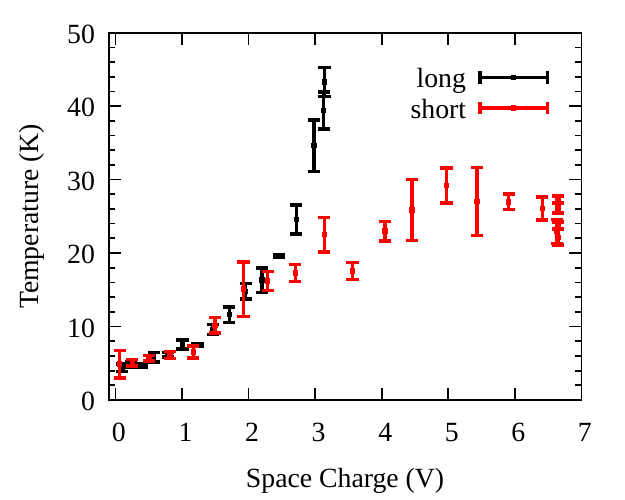} 
    \end{minipage}
    \caption{Evaporative cooling of electrons prepared with SDREVC at $150\,\mathrm{kHz}$ (left panel) or $350\,\mathrm{kHz}$ (right panel). Initial plasma parameters (rightmost points of each dataset) are $N=19\times 10^6$, $r_p=0.81\,\mathrm{mm}$ (left panel) and $N=16\times 10^6$, $r_p=0.57\,\mathrm{mm}$ (right panel).}
    \label{fig:evc}
\end{figure}

Figure~\ref{fig:evc} plots the temperature data for plasma with the initial parameters given in the caption. Data points and error bars represent the mean and standard deviation of 4 measurements; outliers were removed by starting with 5 measurements per bin and removing the one that deviated the most. ``Long'' and ``short'' refer to the well shapes shown in Fig.~\ref{fig:pots}(a). The well bottom is at $50\,\mathrm{V}$ to improve the signal for the temperature diagnostic, which follows with as little manipulation as possible after EVC. 
The plasma is prepared as usual with SDREVC, allowed to cool to $T\approx 30\,\mathrm{K}$, and then evaporatively cooled by reducing the confinement over $300\,\mathrm{ms}$. Different choices of final voltage on the rightmost potential barrier result in different final values of the space charge and temperature. 

For space charge $\phi_0>3\,\mathrm{V}$, EVC does not reduce the temperature of the plasma. For $\phi_0<3\,\mathrm{V}$, the temperature is completely determined by the final value of $\phi_0$. For example, if we want to EVC to $T<15\,\mathrm{K}$, we need to reduce the space charge to $\phi_0<2\,\mathrm{V}$. The value $\phi_0=2\,\mathrm{V}$ corresponds to a range $4.5<N<15\times 10^6$ and $0.3<n<1.2\times 10^8\,\mathrm{cm}^{-3}$ for the four data sets shown. 

The plasma radius $r_p$ increases when $N$ is reduced by EVC. The data for $N$ and $r_p$ (not shown) is well described by the relationship $Nr_p^2=\mathrm{constant}$, which is expected from the conservation of canonical angular momentum\cite{onei:80}. It follows that for a given final temperature, the highest plasma density will be obtained by performing as little EVC as possible. According to our observations, this means reducing the space charge as much as possible prior to EVC (while keeping the plasma compressed). In addition, since the space charge $\phi_0 \propto N/L_p$, more particles may be cooled to a given temperature by making the plasma longer. 

\section{Conclusion}
\label{sec:conclu}
This work offers some perspective on how far the SDR, EVC, and SDREVC methods can be stretched to accommodate extreme or conflicting experimental requirements. In particular, it demonstrates that
\begin{enumerate}
    \item SDR (and SDREVC) require neither a uniform $B$ field nor a pure dipole RW field. Equivalent performance is obtained in a strong mirror field using a RW with only three out of four sectors active.
    \item SDREVC-level reproducibility can be obtained using e-kick instead of EVC.
    \item EVC to the lowest (reported) temperatures ($T\sim 10\,\mathrm{K}$) seems to require reducing the plasma space charge to $\phi_0 < 2\,\mathrm{V}$.
\end{enumerate}

Point (2) may be worth considering for experiments where EVC is not reliable because of electronic noise; slowly opening the confining potential is equivalent to a downward sweep of the plasma bounce frequency, and noisy experiments may have trouble finding a frequency range without resonances. Another possible application comes from the particle-specific nature of the e-kick. So far, SDREVC has not been applied to antiprotons because they can only be compressed by the RW in the presence of electrons, which provide sympathetic cooling\cite{andr:08, aghi:18}. In such a mixed plasma, it seems difficult to control whether electrons or antiprotons are removed by EVC. It is possible that some combination of standard SDREVC for electrons and pulsed SDREVC for the mixed electron-antiproton plasma could stabilize the final number of antiprotons in the presence of fluctuations in beam intensity from ELENA\cite{bart:18}.

Point (3) might be explained by analogy to the heating that occurs when the plasma expands radially: as electrons move to higher radii, $\phi_0$ decreases and some of the potential energy of the electrons becomes kinetic energy. Similarly, one can imagine that kinetic energy is delivered to the plasma during EVC by the recoil from evaporating electrons. For sufficiently large $\phi_0$, this hypothetical mechanism could even cause EVC to heat the plasma more than cool it, as is (marginally) suggested by the data for $\phi_0 > 5\,\mathrm{V}$. If the heating mechanism is as simple as this, then the seemingly arbitrary limit of $\phi_0 \approx 2\,\mathrm{V}$ may turn out to be fairly general.

\end{document}